\documentclass[conference]{IEEEtran}
\IEEEoverridecommandlockouts
\usepackage{cite}
\usepackage{amsmath,amssymb,amsfonts}
\usepackage{algorithmic}
\usepackage{graphicx}
\usepackage{textcomp}
\usepackage{xcolor}
\usepackage{xurl}
\usepackage{booktabs}
\usepackage{color}
\usepackage{url}
\usepackage{multirow}
\usepackage{diagbox}
\usepackage{float}
\usepackage{subfig}
\usepackage{enumitem}
\def\BibTeX{{\rm B\kern-.05em{\sc i\kern-.025em b}\kern-.08em
    T\kern-.1667em\lower.7ex\hbox{E}\kern-.125emX}}
\begin{document}

\title{AudioVMAF: Audio Quality Prediction with VMAF}

% \author{\IEEEauthorblockN{Arijit Biswas}
% \IEEEauthorblockA{\textit{Sound Tech Research} \\
% \textit{Dolby Germany GmbH}\\
% Nürnberg, Germany \\
% arijit.biswas@dolby.com}
% \and
% \IEEEauthorblockN{Harald Mundt}
% \IEEEauthorblockA{\textit{Sound Tech Research} \\
% \textit{Dolby Germany GmbH}\\
% Nürnberg, Germany \\
% harald.mundt@dolby.com}
% }

\author{\IEEEauthorblockN{Arijit Biswas}
\IEEEauthorblockA{\textit{Sound Tech Research} \\
\textit{Dolby Germany GmbH}\\
 arijit.biswas@dolby.com
}
\and
\IEEEauthorblockN{Harald Mundt}
\IEEEauthorblockA{\textit{Sound Tech Research} \\
\textit{Dolby Germany GmbH}\\
 harald.mundt@dolby.com
}
}

\maketitle

\begin{abstract}
Video Multimethod Assessment Fusion (VMAF)~\cite{NetflixTechBlogVMAF},\cite{NetflixTechBlogVMAFbetter},\cite{NetflixTechBlogVMAFJourney} is a popular tool in the industry for measuring coded video quality. In this study, we propose an auditory-inspired frontend in existing VMAF for creating videos of reference and coded spectrograms, and extended VMAF for measuring coded audio quality. We name our system AudioVMAF. We demonstrate that image replication is capable of further enhancing prediction accuracy, especially when band-limited anchors are present. The proposed method significantly outperforms all existing visual quality features repurposed for audio, and even demonstrates a significant overall improvement of 7.8\% and 2.0\% of Pearson and Spearman rank correlation coefficient, respectively, over a dedicated audio quality metric (ViSQOL-v3~\cite{v3}) also inspired from the image domain.
\end{abstract}


\section{Introduction}
\label{section:Introduction}

Vision and audition are the richest sources of sensory data that drive the quality of multimedia experiences. While video storage and transmission command more resources, audio is also a resource hog and is important to consumers. Especially since bandwidth-hungry spatial audio~\cite{NetflixTechBlogAudio} is becoming more pervasive, perceptual audio rate control should be a significant factor in the quality of experience (QoE) optimization. Therefore, sound-quality evaluation tests are critical since they provide the necessary user feedback that drives audio quality improvements. However, subjective listening tests demand a large amount of time and effort. Thus, there is an impetus to develop accurate audio quality assessment models.

In the broader context, perceived multimedia QoE is affected by both perceptual video and audio quality. Quality assessments of audio and video have both been widely researched for decades yet the two areas have been largely mutually independent~\cite{AVQ_survey}. However, the neurosensory systems of the two modalities bear similarities. There are both low-level analogous processing (e.g., masking and multi-scale decomposition) as well as high-level cognitive modeling leading to a useful fusion of the senses~\cite{ERNST2004162}. Furthermore, the modalities themselves bear similarities that make them inter-convertible, e.g., using time-frequency representations of audio (spectrograms) to create a visual format for analysis. So, it is reasonable to consider whether suitable video quality metrics might be adapted for audio quality prediction.

It has been found in subjective joint audio-visual studies~\cite{Bovik_AVQ} that: (1) video modality is relatively more important to QoE than the audio modality; (2) unlike video quality, subjects found it harder to differentiate audio quality (even with audio bitrates chosen to create large degradation), and (3) subjects usually judge the quality of each modality sequentially, before giving an overall rating; suggesting that for a joint audio-visual quality (AVQ) estimation it makes sense to fuse audio and video quality scores using posterior fusion strategy~\cite{beerends1999the},\cite{Bovik_AVQ},\cite{Helard_SeeHear}. Thus, for an AVQ prediction task, observation (1) would indicate that it is reasonable to make use of a video quality metric that is well accepted by the community; observation (2) would hint that one may not need an incoherent (and consequently a complex) system architecture. Therefore, for a coherent system design, it becomes valid to ask the research question, why not derive audio quality from a state-of-the-art video quality metric?

With the above research question, we propose AudioVMAF to measure audio quality with the industry-popular VMAF. We are unaware of techniques that utilize an ``out-of-the-box'' video quality metric as an audio quality metric. We are only aware of: (a) adaptation of a set of classical 2D visual quality indicators for 1D audio signals for measuring audio quality~\cite{Bovik_AVQ}; and (b) adaptation of a 2D image distortion metric (Structural Similarity Index or SSIM~\cite{SSIM}) to a 2D distortion metric (Neurogram Similarity Index Measure or NSIM~\cite{NSIM}) for measuring speech intelligibility and coded speech~\cite{Hines20155} and audio quality~\cite{Hines20156:ViSQOLAudio01},\cite{7940042},\cite{v3} with ViSQOL, where the support vector machine (SVM)~\cite{cortes1995_svm} is trained for the task. In~\cite{Bovik_AVQ}, the 1D variant of 2D visual quality features was evaluated with stereo audio waveform coded with Advanced Audio Coding (AAC) at 8, 32, and 128 kb/s. It is expected that the lowest two bitrates provide an unacceptable audio quality, and differences in decoded audio bandwidth would already serve as a cue for ranking the quality. We believe automatic audio quality assessment becomes a challenge when modern parametric bandwidth extension~\cite{HE-AAC} and parametric stereo coding~\cite{HE-AAC} tools are utilized at lower bitrates. Thus, we evaluated AudioVMAF with modern audio codecs for a wide range of bitrates (i.e., quality). Furthermore, unlike the measures used in \cite{Bovik_AVQ} and ViSQOL, due to the usage of VMAF, we can predict the coded audio quality directly on a 0-100 MUSHRA (Multiple Stimuli with Hidden Reference and Anchor~\cite{MUSHRA}) quality scale; a well-established scale for assessing audio codecs.

The paper is organized as follows. Section~\ref{section:AudioVMAF} describes the AudioVMAF technology. The data used for AudioVMAF evaluation along with the experimental results are given in Section~\ref{section:Results}, and finally, the conclusion is drawn in Section~\ref{section:ConclusionDiscussion}.

\section{AudioVMAF}
\label{section:AudioVMAF}

VMAF predicts the perceptual video quality of a coded video with respect to an uncoded reference video. In VMAF, pixel-level data are pooled to create frame-level image quality measures (Visual Information Fidelity or VIF~\cite{VIFP}, Detail Loss Metric or DLM~\cite{DLM}) modified to cover multiple scales of resolution. Thereafter, different spatial and temporal features are fused using SVM regression to create frame-level quality scores; and finally, consecutive frame scores are pooled to produce the final VMAF score.

\begin{figure}[t]
\begin{center}
  \includegraphics[width=1.0\linewidth]{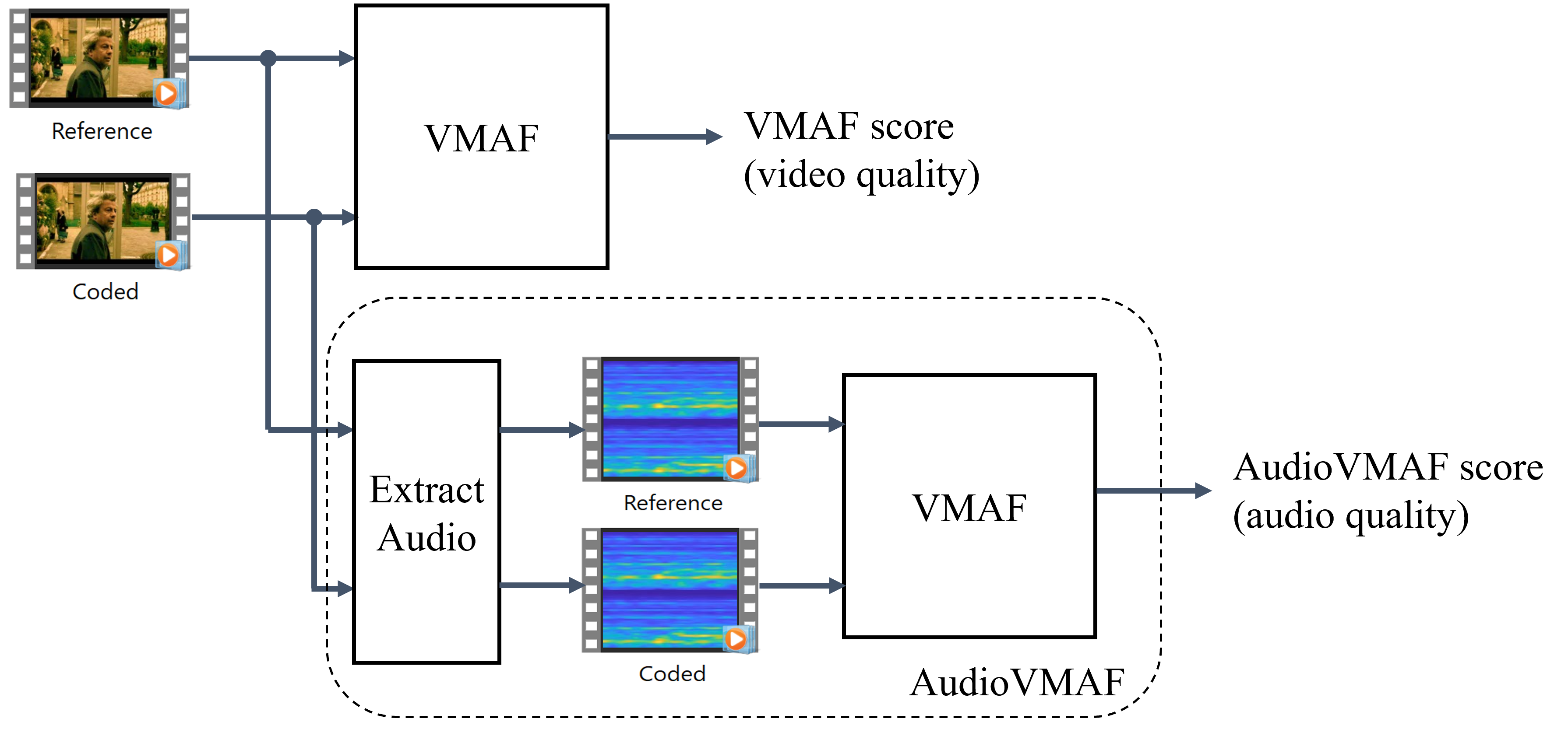} 
  \caption{Utilizing VMAF for both video and audio quality prediction. For AudioVMAF, reference-coded pairs of perceptually motivated spectrogram videos are fed to VMAF.}
  \label{fig:AudioVMAF}
  \end{center}
\end{figure}

For audio quality prediction, reference and coded audio signals are extracted from the corresponding video files (e.g., mp4), and a new set of reference and coded video files are created containing perceptually motivated spectrograms. These video files are then fed into VMAF for computing the audio quality score (Figure~\ref{fig:AudioVMAF}). Our MATLAB-based framework is built around the FFmpeg~\cite{FFmpeg} tool which is used for extracting audio from video, creating videos from images, and for running VMAF. Note that even though the VMAF repository~\cite{VMAFgithub} allows retraining the SVM, in this study, we focused on the proposed auditory-inspired frontend to VMAF. 

Next, we describe how the spectrogram video files are created. Three trivial steps need to be computed before that: (1) audio is extracted from the video files, (2) coded audio is resampled to the sample rate of reference audio, and (3) coded audio is time-aligned with reference audio.

%Technical details
\subsection{Audio to perceptually motivated spectrogram images}
\label{subsection:toSpecVideo}

First, perceptually motivated power spectrograms of the reference and coded audio are computed. For stereo audio, left (L), right (R), and mid-signal (M = 0.5(L+R)) are considered. Note that mid-signal has been considered previously~\cite{7940042},\cite{StereoInSE-NET} for the coded stereo audio quality prediction task. Next, the following perceptually inspired setups are employed: (i) analysis with 80 Equivalent Rectangular Bandwidth (ERB) bands~\cite{moore2012introduction} in the 30~Hz-18~kHz audible frequency range, using FFT with a window length of 42.7 ms and Gammatone filter shape~\cite{moore2012introduction} weighting, (ii) the time stride is aligned with the reference video frame rate (30 fps or 33.3 ms), (iii) the ERB center frequencies were adjusted to the nearest FFT bin center frequencies to avoid sampling issues with the relatively low FFT frequency resolution compared to the narrow Gammatone filter shapes at low frequencies, and (iv) the input audio is calibrated such that $-25$ dB full-scale sine tone corresponds to 85 dB sound pressure level and signals below the threshold-in-quiet~\cite{spanias2006audio} are set to zero.

\subsection{Framing and timing of spectrogram images}
\label{subsection:framing_timing}

Next, we constructed data frames from the reference and coded spectrograms individually as follows. At the video frame rate, 32 spectrogram frames ($\approx$1s) are assembled resulting in 2D arrays of size [80$\times$32] for every audio signal. Then we stacked the 2D arrays for each audio signal on top of each other which, in the case of stereo audio, results in an array of size [240$\times$32] due to the three signals L, R, and M.  The stacking was done for a joint analysis of both individual channels as well as their inter-channel relationships. Finally, for feeding into VMAF, we copy this array as many times as needed to fit into an image of size 480$\times$640 (height$\times$width). The copying (replication) was done to improve the quality prediction accuracy of bandlimited audio (see, results in Section~\ref{subsection:benchmark}). 
%To speed up the processing, the frame computation may be sub-sampled by a factor of 8 without affecting the accuracy.

\subsection{Color and intensity scaling of spectrogram images}
\label{subsection:color_scaling}

The reference and coded image frames are converted to the dB domain with a maximum dynamic range of 70 dB. The dB-domain audio data is then quantized linearly into the [0,255] range. For every frame, the quantized data is used as an index to the HSV (Hue, Saturation, Value) colormap~\cite{hsv_matlab} to create color images in the Portable Network Graphics (PNG) format. This mapping induces a non-monotonic conversion from dB to luma, which is then analyzed by VMAF. Without this mapping from the spectrogram dB-domain to color images using the HSV colormap, we observed significantly worse audio quality prediction performance with monotonic grayscale images (see, results in Section~\ref{subsection:benchmark}).

The design choices of AudioVMAF presented in this section were not informed by the test sets used for its benchmarking. The perceptual front end is similar in spirit to the perceptual frontend used in ViSQOL-v3. The replication method was introduced when we observed that the predicted quality of the bandlimited signals is as high as that of the full-bandwidth signals. The method was found through experiments. Similarly, the HSV color mapping was also found through experiments conducted on a typical MUSHRA listening test (different from the test sets) with 12 excerpts (not present in the test set), involving two variants of a codec (not present in the test set).

\section{Experiments and Results}
\label{section:Results}

\subsection{Test sets}
\label{subsection:testset}

We benchmarked the prediction accuracy of AudioVMAF against subjective listening scores from the Unified Speech and Audio Coding (USAC)~\cite{MPEG-USAC} verification listening tests~\cite{usac_lt} \cite{USAC}. These comprehensive tests contain 24 excerpts coded with USAC, High-Efficiency Advanced Audio Coding (HE-AAC), and Extended Adaptive Multi-Rate – Wideband (AMR-WB+) with bitrates ranging from 8~kb/s mono to 96~kb/s stereo. According to our experience, automatic audio quality assessment becomes a challenge when parametric coding tools are activated. Hence, we evaluated AudioVMAF with modern audio codecs which utilize parametric bandwidth extension and parametric stereo coding tools at lower bitrates. USAC verification listening tests consist of three separate listening tests: mono at low bitrates and stereo at both low and high bitrates. All tests were MUSHRA tests, with a 0-100 quality scale, where a higher score implies better quality. For the details of these MUSHRA listening tests, interested readers are referred to~\cite{usac_lt},\cite{USAC}. For each of the three listening tests, we included all the 24 test excerpts, which consist of 8 speech, 8 music, and 8 mixed excerpts (see, Table 4 in \cite{USAC}).

\subsection{AudioVMAF Benchmarking}
\label{subsection:benchmark}

We compared the accuracy of AudioVMAF against a dedicated audio quality metric which is also inherited from the image domain. We decided to use ViSQOL-v3 (operating in audio mode)~\cite{ViSQOLgithub} because the NSIM distortion metric used in ViSQOL for comparing reference and coded Gammatone spectrograms is inspired by the 2D image distortion metric. Furthermore, it has been reported in \cite{Fraunhofer}, that out of all objective measures designed to evaluate codecs, ViSQOL shows the best correlation with subjective scores and achieves high and stable performance for all content types. Similarly, it was reported in~\cite{smaq_netflix} that overall VISQOL performed very well across five different datasets (see, Table II in~\cite{smaq_netflix}, overall correlation with ViSQOL as the teacher).

\begin{table}[h!]
\setlength\tabcolsep{5pt}
\centering
\caption{Performance of AudioVMAF on USAC verification listening tests. The table shows the correlation coefficients ($R_p$ and $R_s$) between predicted objective scores and subjective (MUSHRA) scores for (a) mono listening test, (b) stereo low-bitrate, and (c) stereo high-bitrate. Note that AudioVMAF (w/ replication) is enabled with the best settings, i.e., both replication and HSV colormap.}

\vspace{-0.4cm}
\begin{center}
\scriptsize{
\begin{tabular}{|l||c|r|c|r|}
\hline
\multirow{2}{*}{\backslashbox{\textbf{Model}}{\textbf{Metric}}}   & \multicolumn{2}{c|}{\textbf{w/ anchors}}   & \multicolumn{2}{c|}{\textbf{w/o anchors}}      \\ \cline{2-5} 
   & $\mathbf{R_p}\uparrow$   & \multicolumn{1}{c|}{$\mathbf{R_s}\uparrow$} & $\mathbf{R_p}\uparrow$   & \multicolumn{1}{c|}{$\mathbf{R_s}\uparrow$} \\ \hline\hline
\textbf{ViSQOL-v3}                  & 0.809             & 0.836             & 0.796             & 0.836   \\ \hline
\textbf{SSIM\textsubscript{1D}}     & 0.240             & 0.232             & 0.697             & 0.623   \\ \hline
\textbf{MS-SSIM\textsubscript{1D}}  & 0.371             & 0.328             & 0.760             & 0.704   \\ \hline
\textbf{VIFP\textsubscript{1D}}     & 0.447             & 0.412             & 0.778             & 0.738   \\ \hline
\textbf{GMSM\textsubscript{1D}}     & 0.130             & 0.175             & 0.677             & 0.584   \\ \hline
\textbf{GMSD\textsubscript{1D}}     & 0.173             & 0.184             & 0.711             & 0.592   \\ \hline
\textbf{AudioVMAF w/o replication}                  & 0.415             & 0.404             & 0.871             & 0.853   \\ \hline
%\textbf{AudioVMAF (CL 6091031)}           & 0.83              & 0.81              & 0.87              & 0.87   \\ \hline
\textbf{AudioVMAF (w/ replication)}           & \textbf{0.870}    & \textbf{0.856}    & \textbf{0.891}    & \textbf{0.885}   \\ \hline
\textbf{AudioVMAF w/o HSV colormap}           & 0.734    & 0.805    & 0.681    & 0.788   \\ \hline
\end{tabular}}
\end{center}
\subfloat[\label{tbl:usac2}]
\newline
\vspace{-0.4cm}
\begin{center}
\scriptsize{
\begin{tabular}{|l||c|r|c|r|}
\hline
\multirow{2}{*}{\backslashbox{\textbf{Model}}{\textbf{Metric}}}   & \multicolumn{2}{c|}{\textbf{w/ anchors}}   & \multicolumn{2}{c|}{\textbf{w/o anchors}}      \\ \cline{2-5} 
   & $\mathbf{R_p}\uparrow$   & \multicolumn{1}{c|}{$\mathbf{R_s}\uparrow$} & $\mathbf{R_p}\uparrow$   & \multicolumn{1}{c|}{$\mathbf{R_s}\uparrow$} \\ \hline\hline
\textbf{ViSQOL-v3}                  & 0.771             & \textbf{0.778}    & 0.689             & 0.694   \\ \hline
\textbf{SSIM\textsubscript{1D}}     & 0.060             & -0.063            & 0.557             & 0.251   \\ \hline
\textbf{MS-SSIM\textsubscript{1D}}  & 0.221             & 0.100             & 0.656             & 0.445   \\ \hline
\textbf{VIFP\textsubscript{1D}}     & 0.213             & 0.158             & 0.339             & 0.313   \\ \hline
\textbf{GMSM\textsubscript{1D}}     & -0.077            & -0.148            & 0.485             & 0.149   \\ \hline
\textbf{GMSD\textsubscript{1D}}     & 0.075             & -0.119            & 0.661             & 0.194   \\ \hline
\textbf{AudioVMAF w/o replication}                  & 0.470             & 0.294             & 0.803             & 0.664   \\ \hline
%\textbf{AudioVMAF (CL 6091031)}     & 0.78              & 0.65              & 0.81              & 0.68   \\ \hline
\textbf{AudioVMAF (w/ replication)}     & \textbf{0.824}    & 0.736             & \textbf{0.815}    & \textbf{0.709}   \\ \hline
\textbf{AudioVMAF w/o HSV colormap}           & 0.669    & 0.714    & 0.551    & 0.656   \\ \hline
\end{tabular}}
\end{center}
\subfloat[\label{tbl:usac3}]
\newline
\vspace{-0.4cm}
\begin{center}
\scriptsize{
\begin{tabular}{|l||c|r|c|r|}
\hline
\multirow{2}{*}{\backslashbox{\textbf{Model}}{\textbf{Metric}}}   & \multicolumn{2}{c|}{\textbf{w/ anchors}}   & \multicolumn{2}{c|}{\textbf{w/o anchors}}      \\ \cline{2-5} 
   & $\mathbf{R_p}\uparrow$   & \multicolumn{1}{c|}{$\mathbf{R_s}\uparrow$} & $\mathbf{R_p}\uparrow$   & \multicolumn{1}{c|}{$\mathbf{R_s}\uparrow$} \\ \hline\hline
\textbf{ViSQOL-v3}                  & 0.823             & 0.904             & 0.769             & 0.852   \\ \hline
\textbf{SSIM\textsubscript{1D}}     & 0.263             & 0.417             & 0.702             & 0.803   \\ \hline
\textbf{MS-SSIM\textsubscript{1D}}  & 0.460             & 0.559             & 0.752             & 0.814   \\ \hline
\textbf{VIFP\textsubscript{1D}}     & 0.389             & 0.517             & 0.332             & 0.581   \\ \hline
\textbf{GMSM\textsubscript{1D}}     & 0.115             & 0.239             & 0.678             & 0.807   \\ \hline
\textbf{GMSD\textsubscript{1D}}     & 0.116             & 0.248             & 0.738             & 0.797   \\ \hline
\textbf{AudioVMAF w/o replication}                  & 0.751             & 0.831             & \textbf{0.834}             & 0.894   \\ \hline
%\textbf{AudioVMAF (CL 6091031)}     & \textbf{0.92}     & \textbf{0.94}     & \textbf{0.84}     & \textbf{0.90}   \\ \hline
\textbf{AudioVMAF (w/ replication)}     & \textbf{0.909}             & \textbf{0.938}             & 0.818             & \textbf{0.898}   \\ \hline
\textbf{AudioVMAF w/o HSV colormap}           & 0.797    & 0.896    & 0.550    & 0.825   \\ \hline
\end{tabular}}
\end{center}
\subfloat[\label{tbl:mono}]

\end{table}

In addition, we benchmark against a related prior research~\cite{Bovik_AVQ}, where 1D variants of the popular video frame/picture quality predictors, e.g., SSIM~\cite{SSIM}, Multi-scale Structural Similarity Index (MS-SSIM)~\cite{MS-SSIM}, Visual Information Fidelity in the pixel domain (VIFP)~\cite{VIFP}, Gradient Magnitude Similarity Mean (GMSM)~\cite{GMSM_GMSD}, and Gradient Magnitude Similarity Deviation (GMSD)~\cite{GMSM_GMSD} were used to predict audio quality. These measures were defined using 1D data windows and are denoted with a subscript 1D in Table I.  We utilize the implementation~\cite{AVQAgithub} provided by the authors~\cite{Bovik_AVQ}. Note that in the implementation, SSIM\textsubscript{1D}, MS-SSIM\textsubscript{1D}, VIFP\textsubscript{1D}, GMSM\textsubscript{1D} are bounded between [0, 1], whereas GMSD\textsubscript{1D} is bounded between [0.687, 1]. ViSQOL-v3 on the other hand is trained to predict the Mean Opinion Score (MOS) (between 1 and 5), but it is observed to be bounded between [1, 4.732]~\cite{InSE-NET},\cite{ViSQOLgithub}. Furthermore, none of the benchmarks are designed for stereo. Internally, ViSQOL-v3 downmixes stereo to a mono mid-signal and then predicts the MOS. Whereas, the 1D visual quality measures (as implemented~\cite{AVQAgithub}) consider only the left channel. However, to make them at least comparable with ViSQOL-v3, we report the results by considering the mono mid signal. For stereo audio, AudioVMAF considers left (L), right (R), and mid-signal (M) by stacking spectrograms of L, R, and M channels in the image. We used the Spearman rank-order correlation coefficient ($R_s$) to measure the prediction monotonicity of the models and the Pearson linear correlation coefficient ($R_p$) to measure the prediction linearity. For both $R_p$ and $R_s$, larger values denote better performance. The prediction accuracy on three listening tests is presented in Table I.       

It can be observed that none of the 1D visual quality features are competitive in predicting the audio quality if the (3.5~kHz and 7~kHz) bandlimited anchors are included. We observed that the quality of the anchors was significantly overestimated (almost as high as the reference quality). Excluding the anchors improves the accuracy, but none of them outperform ViSQOL-v3. Considering all three tests, the top two 1D visual quality features are MS-SSIM and VIFP, indicating the importance of multi-scale modeling and natural statistics-based features also for audio quality prediction. On the contrary, AudioVMAF significantly outperforms 1D visual quality features both with and without the anchors; and performs as well or better than ViSQOL-v3 if anchors are excluded. We further enhanced the performance with anchors by replicating the data (see, Section~\ref{subsection:framing_timing}). Overall, across all three listening tests, we demonstrate a significant improvement ($7.8\%$ and $2\%$ improvement of Pearson's and Spearman's Rank correlation coefficient, respectively) over ViSQOL-v3. We can also observe that if we do not perform the mapping from the spectrogram dB-domain to color images using the HSV colormap (see, Section~\ref{subsection:color_scaling}), the prediction performance of AudioVMAF is significantly degraded. Furthermore, since the predicted AudioVMAF scores are between 0-100, we can also easily compute (without any audio quality scale conversion or mapping) and report the outlier ratios~\cite{yi22b_outlierRatio} for the AudioVMAF with its best settings (i.e., with replication and HSV colormap) enabled: 0.740 (mono test), 0.742 (stereo low bitrate test), and 0.773 (stereo high bitrate test).

\begin{figure}[t]
  \centering
  \includegraphics[width=1.0\linewidth]{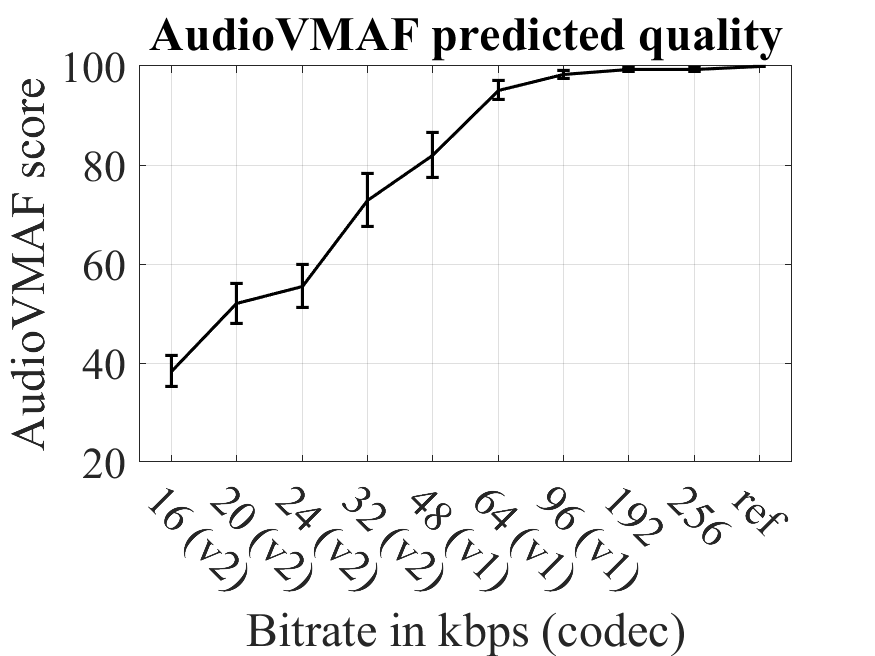} 
  \caption{Scaling of AudioVMAF scores with bitrates: 24 stereo excerpts coded with HE-AAC (v1 or v2) and AAC.}
  \label{fig:quality_scaling}
\end{figure}

Finally, we coded the 24 stereo excerpts for a wide range of bitrates using the (HE)-AAC family of codecs typically used in practice, and demonstrate (Figure~\ref{fig:quality_scaling}) that the mean AudioVMAF score scales with the bitrates. This observation, along with the strong performance in predicting the subjective quality (presented in Table I) makes AudioVMAF also suitable for bitrate laddering (similar to the popular application of VMAF~\cite{VMAF_bitrateladder}).

\section{Conclusion and Discussion}
\label{section:ConclusionDiscussion}

In this paper, we present AudioVMAF: a novel ``out-of-the-box'' VMAF-based coded audio quality prediction model for 48~kHz sample rate. Our contribution can be viewed as a perceptual preprocessing to VMAF. We expanded to predict coded stereo audio quality by stacking spectrograms of left, right, and mid channels in the image. Furthermore, we found that replicating images leads to improved prediction accuracy when bandlimited anchors are included. Interestingly, we also found that the mapping from the spectrogram dB-domain to color images using the HSV colormap is needed for a reasonable prediction performance. In the future, we would like to extend the AudioVMAF for multi-channel coded audio quality prediction, improve the sensitivity of the method at high bitrates, and investigate re-training the SVM with subjective listening test data. Finally, for explaining our findings with AudioVMAF, it is worth understanding the implementation details of VMAF and the dataset that was used for its training. Such insights informed by our observations might trigger ideas for improving also VMAF for image and video quality assessment. 

With this research, we provide a new angle for developing an audio quality metric. The proposed method is a step towards merging coded audio and video quality prediction tasks which may pave a new way for AVQ modeling using a coherent architecture.

\bibliographystyle{IEEEtran}

% Reference to bibliography file.
%\bibliography{IEEEexample}

\end{document}